\documentstyle[12pt]{article}

\begin{document}

\begin{titlepage}
\begin{flushright}

 
\end{flushright}

\begin{center}
{\Large \bf Constraints on heavy $Z^{\prime}$ couplings from 
$\Delta S = 2$ $ B^- \to K^- K^- \pi^+$  decay}\\
\vspace{1cm}
{\large \bf S. Fajfer$^{a,b}$,  P. Singer$^{c}$\\}

{\it a) J. Stefan Institute, Jamova 39, P. O. Box 3000, 1001 Ljubljana, 
Slovenia}
\vspace{.5cm}

{\it b) 
Department of Physics, University of Ljubljana, Jadranska 19, 1000 Ljubljana,
Slovenia}
\vspace{.5cm}

{\it c) Department of Physics, Technion - Israel Institute  of Technology, 
Haifa 32000, Israel}

\end{center}

\vspace{0.25cm}

\centerline{\large \bf ABSTRACT}

\vspace{0.2cm}
The heavy $Z^{\prime}$ boson with family nonuniversal couplings 
can introduce
flavour changing effects. Constraints on nondiagonal $Z^{\prime}$ 
 couplings coming
from the $\mu-e$ conversion in a muonic atom, $K^0 - \bar K^0$ and $B - \bar B$  mixing, $\epsilon$ and 
$\epsilon^{\prime}/ \epsilon$ CP- violating coefficients have been already
established. By using the OPAL upper bound of the branching ratio 
for the
 $B^- \to K^- K^- \pi^+$ decay, we indicate additional constraints 
  on the  $Z^{\prime}$ 
couplings. We comment also on the constraints of $Z^{\prime}$   
couplings coming
from the $b \to d d \bar s$ transition. 
The constraint obtained here from the  
upper bound of the $B^- \to K^- K^- \pi^+$   decay involves a different combination
of couplings than those previously presented, but is much weaker.

\end{titlepage}

In the ongoing  search of physics beyond Standard Model (SM), many 
extensions have been discussed. Recently, the inclusion of 
 an additional heavy 
neutral $Z^\prime$ gauge boson has been analyzed in great 
detail \cite{LP}. Heavy neutral bosons $Z^{\prime}$ are one of 
the better motivated extensions of the Standard Model 
and they appear in grand unified theories, superstring 
theories and theories with large extra dimensions \cite{LP,EL}.  
 A most appealing case is offered by the 
 perturbative heterotic string models with supergravity 
mediated supersymmetry breaking \cite{CL}. In this approach 
the $U(1)^\prime$ and electroweak breaking are 
both driven by a radiative mechanism. 

From the existing direct 
experimental limits of nonobservation of $Z^\prime$ 
at Fermilab or indirect 
limits from precision data at LEP one may deduce 
\cite{LP} that $m_{Z^\prime}> 0$ $ (500$  ${\rm  GeV})$ and the mixing 
angle of $ Z - Z^\prime$ is rather small, $|\theta| \leq 10^{-3}$. 
Indeed, the $Z^\prime$ mass is predicted in many of the 
suggested models to be between $0.5$ and $1$ ${\rm TeV}$ 
[1-3]. 
Moreover, rather stringent limits on the $Z^\prime$ 
couplings have been 
determined from various processes like $\mu - e$ conversion in 
muonic atoms, $K^0 - \bar K^0$ and $B^0 - \bar B^0$ mixing and 
the $\epsilon$ and $\epsilon^{\prime}/\epsilon$ parameters of CP 
- violation \cite{LP}.
On the other hand, arguments have been advanced recently \cite{LP,EL,R,BB} 
that small apparent deviations from SM could be due 
to an extra gauge boson $Z^\prime$ of mass between $400$ and $800$ 
${\rm  GeV}$. These analyses refer to parity violation in atomic cesium 
\cite{R}, various electroweak precision data including 
the coupling of $b \bar b$ pairs \cite{LP,EL} and the forward-backward 
charge asymmetry of high-mass lepton pairs 
produced in $p - \bar p$ collisions \cite{BB}. Thus it 
appears that the existence of a $Z^\prime$ in the $\leq 1 $ ${\rm TeV}$ 
region is still a viable possibility and the search for its 
effect on additional physical processes, or
alternatively, further limitation of its couplings is of 
obvious interest.

   The rare B meson decays are very important in current searches 
   of
physics beyond SM  \cite{AM1}. The study of the $b \to ss \bar d$ 
transition within
SM,  and its extensions like MSSM without and with ${\cal R}$ 
 parity violation \cite{PAUL} 
and two Higgs doublet models \cite{PAUL0}, have indicated that 
$\Delta S= 2 $ rare B
meson decays are very good candidates to search for 
signals of new
physics, since the  $b \to ss \bar d$  transition is very 
small in the SM
having a branching ratio of $10^{-12}$ $-10^{-11}$ \cite{PAUL}. 
Among the discussed $\Delta S= 2 $
decay modes of the B meson \cite{PAUL,FS1}, 
the $B^- \to K^- K^- \pi^+$ decay
provides a good opportunity for investigating physics beyond the
SM. This is due to the fact that long-distance effects in this 
decay
were shown \cite{FS} to be smaller or comparable to the 
short-distance box
diagram, responsible for this decay in SM \cite{PAUL}. 
The OPAL collaboration
has recently set an upper bound on the branching ratio of 
$Br(B^- \to K^- K^- \pi^+)$ $ < 8.8\times10^{-5}$ \cite{OPAL}, which has also been used to obtain
new limits on ${\cal R}$-parity violating couplings in MSSM.
   
   The extreme smallness of $b \to ss \bar d$ in SM leads us to consider in
this note the possibility of the occurence of this transition as a
result of $Z^{\prime}$ exchange. 
Such a tree-diagram mediated by $Z^{\prime}$ , i.e.
$b \bar s \to Z^{\prime}\to  s \bar d$, is indeed allowed in certain 
theoretical
models [1-3]. 
Our aim would then be to 
calculate the predicted rate of the $Z^{\prime}$ -induced 
$b \to ss \bar d$ decay; however, as we describe below, the 
information 
necessary for such a calculation in the form of upper 
limits for the
couplings (or combination of couplings) involved is not available
presently from the previously determined limits on $Z^{\prime}$ couplings. 
Therefore, we shall use our calculation of the $Z^{\prime}$ -induced decay
in conjunction with the OPAL result on  $B^- \to K^- K^- \pi^+$ 
in order to
obtain further constraints on $Z^{\prime}$  couplings.  

  In the analysis of $Z^{\prime}$  couplings we follow the assumptions of [1],
namely the $Z^{\prime}$  gauge coupling is family - nonuniversal 
as suggested by
string models \cite{CL} and as a result there are also 
flavour-changing
couplings. We thus write the $\Delta S = 2$ effective nonleptonic
Lagrangian induced by $Z^{\prime}$ exchange using the same 
notation as in Ref. \cite{LP}:

 \begin{eqnarray}
{\cal L}& =& \frac{ 4 G_F}{{\sqrt 2}} y \lbrace 
B_{21}^{d_L} B_{23}^{d_L} \bar s \gamma_{\mu} d_L 
\bar s \gamma^{\mu} b_L 
+ B_{21}^{d_R} B_{23}^{d_R} \bar s \gamma_{\mu} d_R 
\bar s \gamma^{\mu} b_R \nonumber\\
&+ &B_{21}^{d_L} B_{23}^{d_R} \bar s \gamma_{\mu} d_L 
\bar s \gamma^{\mu} b_R 
+ B_{21}^{d_R} B_{23}^{d_L} \bar s \gamma_{\mu} d_R
\bar s \gamma^{\mu} b_L\rbrace 
\label{e1}
\end{eqnarray}
with
\begin{equation}
y = (\frac{g_2}{g_1})^2 (\rho_1 sin^2 \theta + \rho_2 cos^2 \theta)
\label{e2}
\end{equation}
and $g_1 = e/sin \theta_W$, $g_2$ is the new $U(1)$ 
gauge coupling,  
$\theta$ is a $Z-Z^\prime$ mixing angle and
\begin{equation}
\rho_i = \frac{M_W^2}{M_i^2 cos^2 \theta_W}, 
\label{e3}
\end{equation}
where $M_i$ are the masses of the neutral gauge boson 
eigenstates, $\theta_W$ is the electroweak mixing angle and 
$B_{ij}$ are the unknown  couplings. 

The experimental results on meson mass splitings 
$\Delta m_K$, $\Delta m_B$ and  $\Delta m_{B_s}$ constrain the real parts of 
the squared $Z^\prime$ couplings to quarks \cite{LP} 
\begin{eqnarray}
y |Re[ (B_{12}^{d_{R,L}})^2]| < 10^{-8}, \enspace 
y |Re[ (B_{13}^{d_{R,L}})^2]| < 6 \times 10^{-8}, && \nonumber\\
y |Re[ (B_{23}^{d_{R,L}})^2]| < 2 \times10^{-8}, \enspace 
y |Re[ (B_{12}^{d_{R,L}})^2]| <  10^{-7},
\label{e4}
\end{eqnarray}
while $CP$ violating parameters in the $ K$  meson system bound the imaginary
part of the squared $Z^\prime - d - s$ couplings
\begin{eqnarray}
y |Im[ (B_{12}^{d_{R,L}})^2]| < 8 \times 10^{-11}.  
\label{e5}
\end{eqnarray}
 Other constraints obtained in \cite{EL} contain additional couplings
(i.e. to leptons or with different coefficients) and are not relevant
for the present calculation. It is then more suitable for our purpose to 
rewrite the limits (\ref{e4}) and (\ref{e5}), related to 
$B_{i,j}^{d _{R,L}}$ couplings, in
the following form
\begin{eqnarray}
y |(Re B_{12}^{d_{R,L}})^2 - (Im B_{12}^{d_{R,L}})^2| & < &10^{-8},
\nonumber\\
y|(Re B_{23}^{d_{R,L}})^2 - (Im B_{23}^{d_{R,L}})^2| 
&<& 2\times 10^{-6},\nonumber\\
y|(Re B_{12}^{d_{R,L}}) (Im B_{12}^{d_{R,L}})|& < & 
4\times 10^{-11}. 
 \label{e5a}
\end{eqnarray}
We turn now to the calculation of the $ B^- \to K^- K^- \pi^+$ transition 
via $Z^\prime$ exchange and to explore the limits on 
$Z^\prime$ couplings provided by the existing 
experimental upper limit for it \cite{OPAL}.

The amplitude for $ B^- \to K^- K^- \pi^+$ 
can be obtained  using the effective Lagrangian given in 
(\ref{e1}). For the calculation of the matrix elements of the operators appearing in 
 the effective Lagrangian we use  
  the factorization approximation.  
This requires the knowledge of the 
 matrix elements of the current operators or the density operators. 
 Here we use the standard form factor representation 
 \cite{ALI,WSB1} of the following 
 matrix elements: 
 \begin{eqnarray}
\langle P'(p') | \bar q_j\gamma^{\mu} q_i|P(p)\rangle & = &
F_1(q^2) (p^{\mu} + p'^{\mu} - \frac{m_P^2 - m_{P'}^2}{q^2}
(p^{\mu} - p'^{\mu})) \nonumber\\
&+ & F_0(q^2) \frac{m_P^2 -m_{P'}^2}{q^2}
(p^{\mu} - p'^{\mu}), 
\label{e6}
\end{eqnarray}
where $F_1$ and $F_0$ contain the contribution of vector and scalar
states respectively and $q^2 = (p - p')^2$. Also, $F_1(0) = F_0(0)$
\cite{WSB1}. For these form factors, one usually assumes pole
dominance \cite{WSB1,JURE}
\begin{eqnarray}
F_1(q^2) & = & \frac{F_1(0)}{ 1 - \frac{q^2}{m_V^2}}; \enspace
F_0(q^2)  =  \frac{F_0(0)}{ 1 - \frac{q^2}{m_S^2}}, 
\label{e7}
\end{eqnarray}
where $m_V$, $m_S$ are the masses of lowest lying vector and scalar 
resonances. 
Note that for the transition we consider, the amplitude factorizes in such a way that only 
the matrix elements of the vector currents contribute.
The  relevant parametrs 
are taken from \cite{ALI,WSB2,JURE} $F_0^{BK}(0) =  $ 
$F_1^{BK}(0)=0.38$, $F_0^{K\pi}(0) =  $ 
$F_1^{K\pi}(0)=0.996$. The masses of the 
 meson  poles are $m_{\bar b s} (1^{-}) = 5.41 $ ${\rm GeV}$,
$m_{\bar s b}(0^+)=5.89 $ ${\rm GeV}$, 
$m_{\bar d s} (1^-)= 0.892$ 
${\rm GeV}$ and $m_{\bar d s}(0^+)= 1.43$ ${\rm GeV}$ 
\cite{ALI,JURE}. 
We introduce   $s= (p_B - k_1)^2 $, $t = (p_B - k_2)^2 = $ 
and $ u = (p_B - p_\pi)^2$ and then  calculate  the matrix element 
\begin{eqnarray}
\langle K^-(k_1) K^- (k_2) \pi^+ (p_\pi) |(\bar s \gamma_\mu d) 
(\bar s \gamma^\mu b) | B^-(p_B)\rangle = &&\nonumber\\
F_1^{K\pi}(s) F_1^{BK}(s) [ m_B^2 +  m_K^2 + 2 m_\pi^2 - s - 2 t  
- \frac{m_K^2 - m_\pi^2}{s} (  m_B^2 - m_K^2 )]&& \nonumber\\
+ F_0^{K\pi}(s) F_0^{BK}(s)\frac{m_K^2 - m_\pi^2}{s} 
(  m_B^2 - m_K^2 ) + [ s \leftrightarrow t].  
\label{e8}
\end{eqnarray}
When calculating the rate, we 
denote by $C_{Z^\prime}$ the combination of couplings appearing 
from Eq. (\ref{e1}) in the decay:
\begin{eqnarray}
C_{Z^\prime}= y [B_{21}^{d_{L}}B_{23}^{d_{L}} + 
B_{21}^{d_{R}}B_{23}^{d_{R}} + 
B_{21}^{d_{L}}B_{23}^{d_{R}} + B_{21}^{d_{R}}B_{23}^{d_{L}}].  
\label{e9}
\end{eqnarray}
The numerical calculation gives for the  branching ratio 
\begin{eqnarray}
BR(B^- \to K^- K^- \pi^+) & =&3.5|C_{Z^\prime}|^2. 
\label{e10}
\end{eqnarray}
Combining (\ref{e10}) with OPAL upper bound 
for the $B^- \to K^- K^- \pi^+$ decay of 
$8.8\times10^{-5}$ one finds 
\begin{eqnarray}
|C_{Z^\prime}|^2 < 2.5 \times 10^{-5}. 
\label{e11}
\end{eqnarray}
The limit given in (\ref{e11}) may be rewritten in a form similar to
Eq. (\ref{e5a}) as
\begin{eqnarray}
y^2 \{ (Re B_{21}^{d_{L}} Re B_{23}^{d_{L}} -  
 Im B_{21}^{d_{L}}Im B_{23}^{d_{L}} +
Re B_{21}^{d_{L}} Re B_{23}^{d_{R}} -  
 Im B_{21}^{d_{L}}Im B_{23}^{d_{R}}  & + &\nonumber\\
Re B_{21}^{d_{R}} Re B_{23}^{d_{L}} -  
 Im B_{21}^{d_{R}}Im B_{23}^{d_{L}} +
Re B_{21}^{d_{R}} Re B_{23}^{d_{R}} -  
 Im B_{21}^{d_{R}}Im B_{23}^{d_{R}} )^2 & + &\nonumber\\
  (Re B_{21}^{d_{L}} Im B_{23}^{d_{L}} + 
Im B_{21}^{d_{L}}Re B_{23}^{d_{L}} +
Re B_{21}^{d_{L}} Im B_{23}^{d_{R}} + 
Im B_{21}^{d_{L}}Re B_{23}^{d_{R}}  & + &\nonumber\\
Re B_{21}^{d_{R}} Im B_{23}^{d_{L}} + 
Im B_{21}^{d_{R}}Re B_{23}^{d_{L}} +
Re B_{21}^{d_{R}} Im B_{23}^{d_{R}} + 
Im B_{21}^{d_{R}}Re B_{23}^{d_{R}})^2\} &  &\nonumber\\
 < 2.5 \times10^{-5}.
\label{e13}
\end{eqnarray}
Inspection of the left hand side of (\ref{e13}) reveals that the necessary
information needed to present an upper limit for the $Z^{\prime}$ 
induced 
 $ B^- \to K^- K^- \pi^+$ decay cannot be derived from the relations 
 summarized in  (\ref{e5a}), unless we assume some of the 
couplings
to vanish. Hence, Eq.(\ref{e13}) should be viewed as an additional constraint
on the Z' couplings, which is not obtainable from the previously
considered processes.The existing upper limit on the  
$ B^- \to K^- K^- \pi^+$ 
branching ratio is rather poor at present and does not allow yet
to obtain a constraint on couplings in the same range as in 
(\ref{e5a}).

Now we briefly comment on possible constraints arising from 
the $b \to d d \bar s$ decay. 
The effective Lagrangian contributing to this transition is 
\begin{eqnarray}
{\cal L}& =& \frac{ 4 G_F}{{\sqrt 2}} y \lbrace 
B_{12}^{d_L} B_{13}^{d_L} \bar d \gamma_{\mu} s_L 
\bar d \gamma^{\mu} b_L 
+ B_{12}^{d_R} B_{13}^{d_R} \bar d \gamma_{\mu} s_R 
\bar d \gamma^{\mu} b_R \nonumber\\
&+ &B_{12}^{d_L} B_{13}^{d_R} \bar d \gamma_{\mu} s_L 
\bar d \gamma^{\mu} b_R 
+ B_{12}^{d_R} B_{13}^{d_L} \bar d \gamma_{\mu} s_R
\bar d \gamma^{\mu} b_L\rbrace .
\label{e1d}
\end{eqnarray} 
Instead of the combination of couplings given in
(\ref{e9}) an experimental
bound on the decay rate of $B^- \to \pi^- \pi^- K^+$
will limit the 
following combination 
$y [B_{12}^{d_{L}}B_{13}^{d_{L}} + 
B_{12}^{d_{R}}B_{13}^{d_{R}} + $ $ 
B_{12}^{d_{L}}B_{13}^{d_{R}} + B_{12}^{d_{R}}B_{13}^{d_{L}}]$,  
which can also be expressed in a form similar to (\ref{e13}).

In concluding, we remark that the rare decays $B^- \to K^- K^- \pi^+$
and $B^- \to \pi^- \pi^- K^+$ 
can provide additional information on the couplings
appearing in the $Z^\prime$  induced nonleptonic Lagrangian, 
which is
complementary to that obtained from mass differences and 
CP-violating
parameters. The new relation obtained here is given in 
Eq.(\ref{e13}). Its limit
is much less stringent than those in (\ref{e5a}), 
since the considered rare
decays are less advantageous presently than the mass 
differences in
obtaining limits for couplings.

This work has been supported in part by the Ministry of 
Science of the Republic of Slovenia (SF) and by the Fund for Promotion 
of Research at the Technion (PS). One of us (PS) is grateful for the
hospitality of the Physics Department of University College London during 
summer 2001, when this work was completed.



\end{document}